\documentclass[twocolumn]{article}

\usepackage[utf8]{inputenc}
\usepackage[english]{babel}
\usepackage{tikz}
\usepackage{amsmath}
\usepackage[lofdepth,lotdepth]{subfig}
\usepackage{listings}
\usepackage{hyperref}
\usepackage{tabularx}
\usepackage{multicol}
\usepackage{soul}

\usepackage{smartdiagram}
\usepackage{placeins}
\begin{document}

\begin{titlepage}

	\title{
    \vspace{1cm}
		\textbf{A Comprehensive Review and Taxonomy of Audio-Visual Synchronization Techniques for Realistic Speech Animation \\ 
		\vspace{0.2in}
		}
  
            \normalsize{Jose Geraldo Fernandes$^*$, \ Sinval \ Nascimento$^+$, \ Daniel \ Dominguete$^+$, \ André \ Oliveira$^+$, \\ Lucas \ Rotsen$^+$, \ Gabriel \ Souza$^+$, \ David \ Brochero$^*$, \ Luiz \ Facury$^*$, \ Mateus \ Vilela$^*$, \\ \ Hebert \ Costa$^*$, \ Frederico \ Coelho$^*$, \ Antônio \ P. \ Braga$^*$} \\
		\vspace*{0.3in}
  
$^*$ Computational \ Intelligence \ Laboratory. \ Department \ of \ Electronic \ Engineering, \\ School \ of \ Engineering. \ Universidade \ Federal \ de \ Minas \ Gerais - \  Brazil \\
$^+$ OMNILOGIC \ Artificial \  Intelligence \ and \ Machine \ Learning \\
Belo \ Horizonte \ - \ Brazil}
\end{titlepage}
\date{}
\maketitle
%\tableofcontents
%\pagebreak
%\setcounter{page}{1}

%================================= ET 1 =========================================
\section{Introduction}\label{sec:introducao}

In several applications, there is a particular emphasis on synchronizing audio with visuals. Examples include creating graphic animations for films or games, where the goal is to synchronize audio with the speaker's facial movements, translating audio from movies into different languages, and even developing applications for the metaverse. Another application, which is the focus of our research, involves developing virtual assistants capable of engaging in realistic real-time conversations with customers. In such an application, when a customer asks a question, it must generate an audio response, synchronize it with an avatar image, and provide a timely conversational reply. For instance, \textit{Audio2Face} (A2F) \cite{karras2017audio} has demonstrated the ability to generate 3D facial animation from audio inputs by learning a mapping function between the input speech audio and key points on the model's face.

In the rapidly evolving field of audio-visual synchronization, achieving realistic facial animations from audio inputs is a critical area of innovation. This review explores a range of methodologies developed to tackle this challenge, highlighting both generative models that create animations from scratch and adaptive models that modify existing footage. Addressing key challenges such as model training costs, dataset availability, and silent moment distributions in audio data, this study presents innovative solutions that enhance performance and realism. Overall, it advances the capabilities of virtual assistants, gaming, and interactive digital media through state-of-the-art techniques and methodologies.

Finally, there are various techniques on the domain of audio and video synchronization for speech animation. However existing literature lacked a clear taxonomy to organize these methods. Consequently, during the research a new taxonomy was developed. The proposed taxonomy categorizes methods based on their logistical aspects while remaining independent of specific model structures or technological details.

\section{Related Work and Proposed Taxonomy}

  As mentioned before in the literature there are several methods to perform synchronization between audio and video, animating generic speech.  
  %\st{In other words, the problem is to synthesize a person video corresponding to an input speech. As we researched existing works in the literature, we realized the need to define a taxonomy to categorize these approaches. Our taxonomy is specific to the logistic dimension of each method, keeping the independence of model structure and other aspects of technological nature.}
  During the literature review process, a new taxonomy needed to be defined in a way to be specific to the logistic dimension of each method, keeping the independence of model structure and other aspects of technological nature.

\begin{figure*}[t!]
    \centering
    \includegraphics[width=1.0\linewidth]{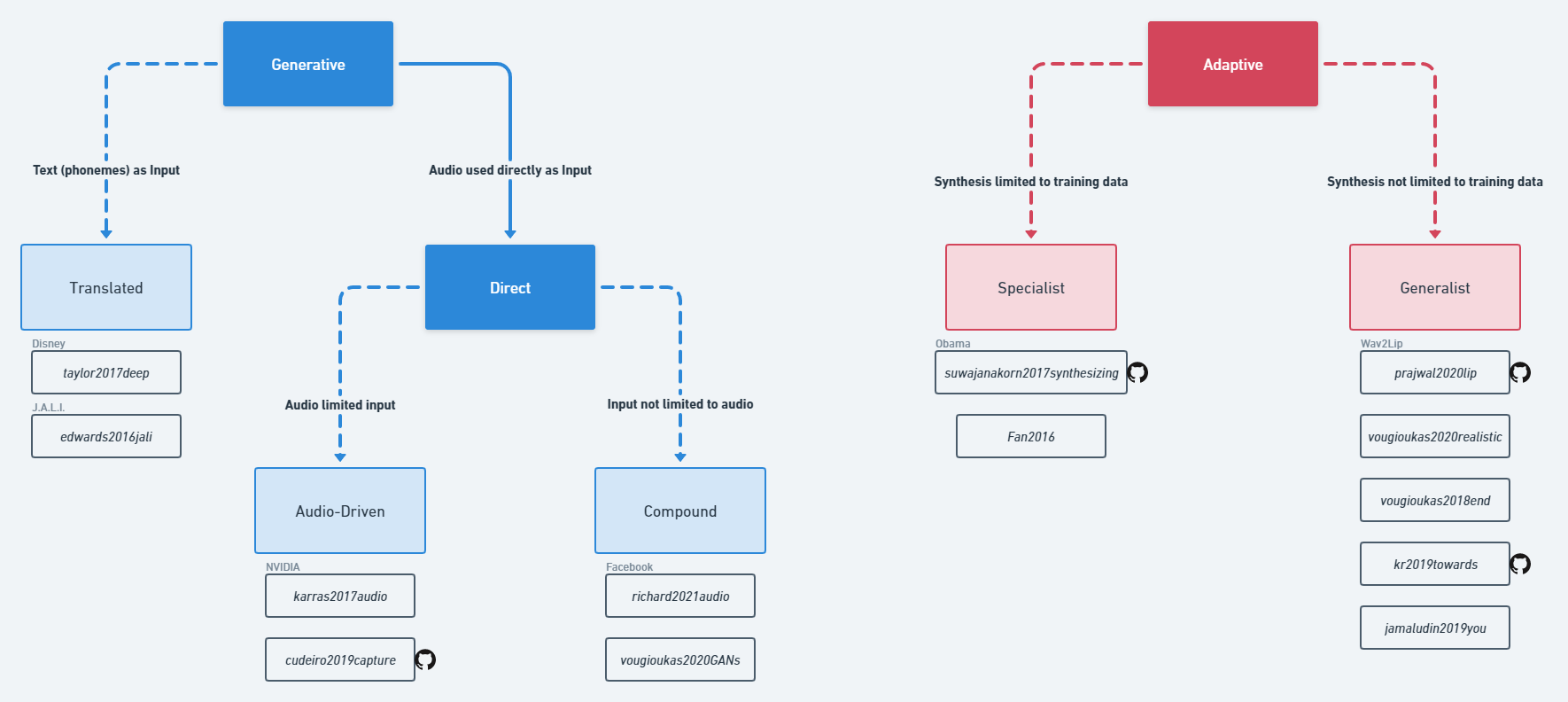}
    \caption{Classification tree of proposed taxonomy.}
    \label{fig:map}
  \end{figure*}

  An overview of proposed taxonomy can be seen in Figure \ref{fig:map}. Through this section, each branch is explained, the group of papers that justify it is cited and %\st{its consequence in a possible application are mentioned} 
  the potential implications of each branch in real-world applications are discussed. Works with repositories are also indicated in Figure \ref{fig:map}. As a general overview, in the first fork,  generative and adaptive approaches are distinguished. The first one refers to a full end-to-end animation synthesizer, commonly working with 3D vertex coordinates of a face model \cite{karras2017audio,richard2021audio}. From a default matrix, these approaches learn the mapping in which each vertex articulates from the input speech, later distinguished in format by our taxonomy. On the other hand, the adaptive branch is specified as a post-production technique. From a finished animation or video, the effort is to synchronize it with the target speech, like the lip-syncing problem  \cite{prajwal2020lip,suwajanakorn2017synthesizing}.

At the end of this section a comparative table (\autoref{table:comparative}) regarding the analyzed papers is provided. The table classifies the papers on type, if it is Multispeaker or not, Input, Output and implementation Difficulty.

\subsection{Generative Translated Branch}
  The first branch to be analyzed is the "Generative Translated Branch". It is Generative, so the animation is synthesized on a model rig and not adapted on top of an existing video or image. Beyond that, the audio tracks are translated to text inputs that represents the phonemes.
  
  %\st{On this branch, we can highlight that t}
  These models are usually independent of speaker, style or language, %\st{mostly thanks to the use of phonemes to distinguish between different emotions and mouth movements} 
  largely due to their utilization of phonemes to differentiate between distinct emotions and mouth movements. While some of the papers like \cite{edwards2016jali} have a more "study-based" approach with the use of FACS (\textit{Facial Action Coding System}) on the model, other studies like \cite{taylor2017deep} propose a model that is capable of real-time synthesis and "character-interchangeable" animation.
  
  It is important to note that these models are limited to mouth animation exclusively, with the rest of the face being animated by professionals when required. Also, the use of text on training and synthesis instead of audio contributes to make the datasets lighter as well, however, the use of such phonemes requires a voice decoder to translate the audio files to text files.

    \subsubsection{A Deep Learning Approach For Generalized Speech Animation \cite{taylor2017deep}}\label{sec:deep learning approach}
    The focus of this study is to develop a speech animation model that can be easily integrated with existing pipelines, generating animations in real time that are editable by professional animators without any trouble and can be switched to other characters. As mentioned, the use of phonemes makes the model speaker independent and the datasets lighter, but at the same time it makes necessary to use external voice decoders in order to translate from audio to phonemes in text form.

    The efficacy of this approach is demonstrated through a diverse array of animation clips featuring different characters and voices, spanning various scenarios such as singing performances and foreign language input. Notably, the system can also generate real-time speech animation directly from user input, further enhancing its usability and accessibility.

    Overall, this work represents a significant advancement in the field of speech animation, offering a powerful and versatile toolset for creating compelling and lifelike animated content across a wide range of applications.
    %\textcolor{gray}{A Deep Learning Approach For Generalized Speech Animation}

%    \fotenotetext[1]{Scale from one star meaning easy, to five stars meaning very hard}

\subsubsection{JALI: An Animator-Centric Viseme Model for Expressive Lip Synchronization \cite{edwards2016jali}}\label{sec:JALI}
Similar to the previously mentioned work, the JALI model generates animations that are editable by animators through the use of phonemes in the form of audio and its respective transcription. The facial rigs used on the study are based on the Facial Action Coding System (FACS), meaning that the visemes on the face are built on Action Units (AU), thus simulating muscular and bone-based movements through lip and jaw variations. The use of such system makes need of a compatible facial rig and external tools to translate and align the phonemes with the audio track.

The paper provides a compelling array of animation clips demonstrating the effectiveness of the proposed system, comparing its outputs against those from performance capture techniques and existing procedural animation methods. Additionally, a brief user study is conducted to evaluate the system's performance and user experience, further validating its efficacy and utility in practical applications within the field of facial animation.

%\textcolor{gray}{JALI: An Animator-Centric Viseme Model for Expressive Lip Synchronization}

\subsection{Generative Direct Compound Branch}

  %\st{The Generative Direct Compound branch is, unlike the other generative models, makes them, in general,  more expensive to train, since it requires more data than usual. On the other hand, these algorithms tend to achieve better results than other generative models, since they benefit from the correlation between the inputs, resulting, many times, in a more realistic output.}
  The Generative Direct Compound branch, unlike other generative models, generally demands a higher training cost due to its need for a larger dataset. However, these algorithms tend to achieve better results than other generative models, since they benefit from the correlation between the inputs, resulting many times in a more realistic output.

\subsubsection{Audio- and Gaze-driven Facial Animation of Codec Avatars \cite{richard2021audio}}\label{sec:facebook}
      Facebook proposed an extremely solid and realistic Generative Direct Compound model, in which raw audio and gaze direction are given as input to the model, and the resulting output is able to animate both the upper and lower parts of the face, including mouth, nose, eyes and eyebrows. This multimodal fusion approach that dynamically determines which sensor encoding should animate specific parts of the face at any given time. This dynamic allocation of animation resources allows for generating full-face motion. The authors %\st{of the paper}
      claim that the input data is relatively easy to obtain, since any Virtual Reality glass can capture them. In spite of the high cost to train this model, it can easily outperform its competitors from the Generative branch in terms of realism.

      %\st{There is a supplemental video accompanying the paper}
      The authors provide a video that demonstrates the system's capability to generate highly realistic and expressive facial animations, showcasing a significant advancement in the field compared to previous approaches. The ability to animate Codec Avatars in real-time opens up new possibilities for immersive virtual reality experiences, enhancing the realism and emotional engagement of virtual interactions.

\subsubsection{Realistic Speech-Driven Facial Animation with GANs \cite{vougioukas2020GANs}}\label{sec:GAN}
    This paper proposes an end-to-end system that generates videos of a talking head, using only a still image and an audio clip. The generated videos present lip synchronization and natural face expressions with blinks and eyebrow movements.

    At a high level, the model's architecture is a GAN with one generator and three discriminators. The generator network has an encoder-decoder structure with a latent representation that is made up of three components, which account for the speaker identity, audio content and spontaneous facial expressions. These components are generated by different modules and combined to form an embedding which can be transformed into a frame by the decoding network. In sequence, the generated frames feed the discriminators, which are focused on three aspects: first, image detail, including identity and sharpness, second, plausible face expressions and finally, audio-visual correspondence.
    
    One main aspect in \cite{vougioukas2020GANs} are the proposed evaluation techniques, like a blink detector and audio-visual sync metrics, which guide the GAN during the training. Beyond the quantitative evaluation, it's also implemented a user study that evaluates the realism of the generated videos through an online Turing test, achieving the excellent result of 52\% of the videos labeling correctly. It's important to highlight that the trained model generalizes well to any voice and face, mainly because it was training over rich data: videos of multiple speakers uttering rich phonetically phrases.

\subsection{Generative Direct Audio Driven Branch}
  
  %\st{Generative methods that take simply audio data as input are here named as Generative Direct Audio Driven. Strategies going through this approach provide a direct translation of voice audio into animated facial movements. The input audio might be natural or TTS synthesized and the general goal is to train a model able to make the real time facial animations out of speakers voice with the lowest latency possible.}
  Generative methods that exclusively utilize audio data as input are hereby termed Generative Direct Audio Driven methods. Strategies following this approach aim to directly translate voice audio into animated facial movements. The input audio could either be natural or text-to-speech synthesized. The overarching objective is to train a model capable of generating real-time facial animations from a speaker's voice with minimal latency.
  
    \subsubsection{Audio-Driven Facial Animation by Joint End-to-End Learning of Pose and Emotion \cite{karras2017audio}}\label{sec:NVIDIA}
      %\st{Aiming in game realistic dialogues, this study present a technique for driving 3D facial animation by audio input in real time.}
      In pursuit of lifelike dialogues within gaming contexts, this study introduces a technique for real-time 3D facial animation driven by audio input. To achieve %\st{their so claimed}
      low latency, a DNN maps input wave forms to per-frame 3D vertex coordinates positions of a fixed topology face model. Simultaneously, the network also learns a compact latent code that captures variations in facial expression not directly attributable to the audio signal. This latent code serves as an intuitive control mechanism for manipulating the emotional state of the animated face in real-time, essentially acting as a "face puppet" controlled by the audio input.

      The network is trained using 3 to 5 minutes of high-quality animation data obtained through traditional performance capture methods based on vision. While the primary focus is on modeling the speaking style of a single actor, the trained model demonstrates reasonable performance even when driven by audio from different speakers with varying gender, accent, or language. This generalization capability is validated through a user study, showcasing the versatility and robustness of the proposed approach.
      
      The results and capabilities of this technique have broad applications across various domains, including in-game dialogue systems, cost-effective localization efforts, virtual reality avatars, and telepresence solutions. By enabling dynamic and emotive facial animations synchronized with audio input in real time, the proposed method enhances the immersive and interactive experiences in virtual environments, making it a valuable tool for developers and content creators in the digital media and entertainment industries.

    \subsubsection{Capture, Learning and Synthesis of 3D Speaking Styles \cite{cudeiro2019}}\label{sec:cudeiro}
    %\st{The article claims that realistic facial animation is an unsolved problem due to the lack of available 3D data sets.}
    The article asserts that achieving realistic facial animation remains an unresolved challenge primarily because of the scarcity of accessible 3D datasets. Therefore, the authors propose data set with approximately 29 minutes of data recorded from 12 different speakers with a 4D scan. A neural network model called VOCA (Voice Operated Character Animation) is, than, trained on the proposed data set. The model is capable of generating a 3D Character Animation with only audio as input, being able to animate the entire bottom of the head, including neck, chin, and mouth. Since VOCA cannot animate the upper face, the output is limited although the results of the botton part can be impressive. Also it is possible to adjust shape of the face and style of speech. By conditioning on subject labels during training, the model learns to produce a variety of realistic speaking styles, further enhancing its versatility, being able to animate any language, speaker or speech.

    One of the key features of VOCA is its provision of animator controls, allowing users to modify speaking style, identity-dependent facial shapes, and pose parameters such as head, jaw, and eyeball rotations during the animation process. Importantly, VOCA stands out as the only realistic 3D facial animation model that can be readily applied to unseen subjects without requiring retargeting, making it highly suitable for applications such as in-game videos, virtual reality avatars, and scenarios where the speaker, speech content, or language is unknown in advance.
    
    %\st{To facilitate further research and development in this domain, the authors have made both the dataset and the trained VOCA model publicly available. This initiative not only contributes to advancing the state-of-the-art in audio-driven facial animation but also fosters collaboration and innovation within the research community.}
    To propel ongoing research and innovation in this field, the authors have made both the dataset and the trained VOCA model openly accessible. This endeavor not only enhances the advancement of audio-driven facial animation techniques but also cultivates collaboration and creativity within the research community.
    
\subsection{Adaptive Specialist Branch}

%\st{As other adaptive ones, models of this type change previous existing footage overwriting characters images to adapt face expressions through time combining original video with new audio. As other specialists these can only produce results of one character known during training.}
Similar to other adaptive models, these models modify pre-existing footage by replacing character images to synchronize facial expressions with new audio inputs over time. However, like other specialized systems, these models are typically trained to produce results for a specific character known during the training process.

    \subsubsection{A deep bidirectional LSTM approach for video-realistic talking head \cite{Fan2016}}\label{sec:LSTM}
        %\st{The proposed technique is based on recurrent networks to superpose mouth region images onto talking faces. The input data requires frontal view recorded from single subject video, accompanied by the corresponding audio or text transcript. Using this pair, the model is trained to learn the sequence mapping from the audio, or text, domain to the visual domain.}
        The proposed technique relies on recurrent networks to overlay mouth region images onto speaking faces. The input data necessitates a frontal view captured from a single subject video, along with the corresponding audio or text transcript. By leveraging this paired information, the model is trained to understand the sequential mapping from the audio or text domain to the visual domain. This trained model can then accurately predict the active appearance model (AAM) parameter trajectories for lower face animation when presented with unseen speech audio, whether it's original recorded speech or synthesized via (TTS) systems. To further enhance the realism of the talking head generated by the model, a trajectory tiling method is employed. This method uses the predicted AAM trajectory as a guide to select smooth, real sample image sequences from the recorded database. These selected lower face image sequences are then combined with a background face video of the same subject, resulting in a final video-realistic talking head.

        %\st{The experimental results demonstrate that the proposed approach surpasses the existing Hidden Markov Model based methods in both objective and subjective evaluations as is the older paper of our review.}
        The experimental results conclusively illustrate that the proposed approach outperforms existing Hidden Markov Model-based methods in both objective and subjective evaluations. This indicates the effectiveness and superiority of the approach in generating highly realistic and convincing video-realistic talking heads, making it a significant advancement in the field of audio-visual modeling and animation.

    \subsubsection{Synthesizing Obama: Learning Lip Sync from Audio \cite{suwajanakorn2017synthesizing}}\label{sec:obama}
           %\st{This work proposes a model that, given an audio track of Barack Obama, synthesizes a high quality video of him speaking with accurate lip sync. The output video is basically a target video that is modified on the mouth area, but preserves the remaining movements and scenario, and is important to note that the model is made based on only one speaker.}
           This study introduces a model designed to generate a high-quality video of Barack Obama speaking with precise lip sync, utilizing an audio track of his voice. The resulting video maintains the original movements and scenario, with modifications limited to the mouth area. It's crucial to emphasize that the model is trained on only a single speaker.

           The output of this approach is described as photorealistic, indicating that the synthesized video of President Obama speaking appears highly realistic and natural. %\st{This achievement highlights the advancement of deep learning techniques in audio-visual synthesis, enabling the creation of convincing and compelling visual content that seamlessly blends audio and visual elements.}
           This accomplishment underscores the progress of deep learning techniques in audio-visual synthesis, facilitating the development of visually persuasive and captivating content that seamlessly integrates audio and visual components.

\subsection{Adaptive Generalist Branch}

%\st{The Adaptive Generalist branch is another approach to solving the lip-syncing problem while keeping the model general to the identities of the targets. While this can be viewed as a technological improvement, one has to recognize that it also demands larger and more diverse datasets, for instance.}
The Adaptive Generalist branch offers an alternative solution to the lip-syncing challenge while maintaining the model's adaptability across various target identities. While this approach represents a technological advancement, it requires larger and more diverse datasets to achieve optimal performance.

|%\st{Another obstacle that such models have to deal with is model complexity. While the usage of GANs is typical to deal with the lip-syncing problem, like in \cite{kr2019towards}, to reconstruct each frame of the input, this approach struggled to show a reliable result in the final video output. Therefore, to treat this, the authors in \cite{prajwal2020lip} design a parallel and pre-trained module, focusing solely on the lip reconstruction, to joint its output to the cost function.}
Another challenge that these models encounter is the complexity of the model itself. While employing Generative Adversarial Networks (GANs) is common practice to address the lip-syncing issue, as seen in \cite{kr2019towards}, where each frame of the input is reconstructed, this method often fails to produce consistently reliable results in the final video output. To address this limitation, the authors in \cite{prajwal2020lip} introduce a parallel and pre-trained module specifically dedicated to lip reconstruction. This module's output is then integrated into the overall cost function to enhance performance.

\subsubsection{Towards Automatic Face-to-Face Translation \cite{kr2019towards}}\label{sec:face-face}

The authors present an application of GANs to automatically synchronize the lip movement of a person's video to another target audio. The model concatenates both faces in video and audio in an encoder-decoder structure and trains it to reconstruct each frame correctly.

The proposed approach establishes an automated pipeline for this task, showcasing its practical utility in various real-world applications. Initially, a speech-to-speech translation system is constructed by integrating multiple existing modules from the fields of speech and language processing. Building upon this foundation, the authors introduce a novel visual module called LipGAN, designed specifically for generating lifelike talking faces that align with the translated audio.

To assess the effectiveness of LipGAN, quantitative evaluations are conducted using the standard LRW \cite{chung2017lip} test set, demonstrating its superior performance across all standard metrics compared to existing approaches. Furthermore, the pipeline undergoes rigorous human evaluations, revealing a significant enhancement in the overall user experience when consuming and interacting with multimodal content across different languages.

%\st{The authors contribute to the research community by making their code, models, and a demo video publicly available, encouraging further exploration and development in the domain of audio-visual translation systems. Overall, the paper introduces an innovative solution that bridges the gap between language translation and realistic visual representation, paving the way for enhanced cross-lingual communication and interaction in digital platforms.}
The authors make a significant contribution to the research community by openly sharing their code, models, and a demo video, thereby fostering continued exploration and advancement in the field of audio-visual translation systems. In essence, the paper presents an inventive solution that effectively connects language translation with lifelike visual representation, laying the foundation for improved cross-lingual communication and interaction across digital platforms.

\subsubsection{A Lip Sync Expert Is All You Need for Speech to Lip Generation In the Wild \cite{prajwal2020lip}}\label{sec:lip-sync}

The authors present an improvement to their previous lip-sync model \cite{kr2019towards} to deal with problems of an unconstrained input, like arbitrary person's identities, unseen in the training data, and position and movement. The proposal is a new pre-trained module, the Lip-Sync Expert, to combine its output with the cost function.

Moreover, the study introduces new and rigorous evaluation benchmarks and metrics tailored specifically for measuring lip synchronization in unconstrained videos. Through extensive quantitative evaluations on these challenging benchmarks, the researchers demonstrate the remarkable lip-sync accuracy achieved by their Wav2Lip model, showcasing results that are almost on par with real synced videos.

To further illustrate the impact of their model, a demo video is provided, clearly showcasing the substantial improvements in lip-syncing accuracy. Additionally, the researchers make their code, trained models, and evaluation benchmarks publicly available on their website, encouraging further exploration and development in the field of lip-syncing technology for talking face videos.

\begin{table*}[h!]
\caption{Comparative Table, I: Type, Generative/Adaptative; II: Multispeaker; III: Input type, Audio/Text/Video/Other; IV: Output type, Animation/Video; V: Target synthesis, Mouth/Face; VI: Speaker independent; VII: Implementation Difficulty; VIII: Open source }
\label{table:comparative}
\begin{tabularx}{1.0\textwidth} { 
      | >{\raggedright\arraybackslash}X
      | >{\raggedleft\arraybackslash}X
      | >{\raggedright\arraybackslash}X
      | >{\raggedright\arraybackslash}X
      | >{\raggedright\arraybackslash}X 
      | >{\raggedright\arraybackslash}X 
      | >{\raggedright\arraybackslash}X 
      | >{\raggedright\arraybackslash}X 
      | >{\raggedright\arraybackslash}X | }
     \hline
         \textbf{Paper} &
         \textbf{I} &
         \textbf{II} &
         \textbf{III} &
         \textbf{IV} &
         \textbf{V} &
         \textbf{VI} &
         \textbf{VII} &
         \textbf{VIII}\\
     \hline
        \scriptsize \cite{taylor2017deep} &
        G &
        Yes &
        T &
        A &
        M &
        Y &
        $\star$ &
        N \\
     \hline
        \scriptsize \cite{edwards2016jali} &
        G &
        Yes &
        AT &
        A &
        M &
        Y &
        $\star\star$ &
        N \\
     \hline
        \scriptsize \cite{richard2021audio} &
        G &
        Yes &
        AO &
        A &
        F &
        Y &
        $\star\star$ &
        N \\
     \hline
        \scriptsize  \cite{vougioukas2020GANs} &
        G &
        Yes &
        AIO &
        V &
        F &
        Y &
        $\star\star$ &
        N \\
     \hline
        \scriptsize \cite{karras2017audio} &
        G &
        Yes &
        A &
        A &
        F &
        Y &
        $\star\star\star$ &
        N \\
     \hline
        \scriptsize \cite{cudeiro2019} &
        G &
        Yes &
        A &
        A &
        M &
        Y &
        $\star$ &
        Y \\
     \hline
        \scriptsize \cite{Fan2016} &
        A &
        No &
        AT &
        V &
        M &
        N &
        $\star$ &
        N \\
     \hline
        \scriptsize \cite{suwajanakorn2017synthesizing} &
        A &
        No &
        AV &
        AV &
        M &
        N &
        $\star\star\star$ &
        Y \\
     \hline
        \scriptsize \cite{kr2019towards} &
        A &
        Yes &
        A &
        V &
        F &
        N &
        $\star\star$ &
        Y \\
     \hline
        \scriptsize \cite{prajwal2020lip} &
        A &
        Yes &
        A &
        V &
        F &
        Y &
        $\star\star$ &
        Y \\
     \hline
     
\end{tabularx}
\end{table*}

\section{Conclusions}\label{sec:conclusao}

The review presented highlights the diverse methodologies and innovative approaches developed to achieve audio-visual synchronization, particularly focusing on generating realistic facial animations from audio inputs. These methods span from generative models that create animations from scratch to adaptive models that modify existing footage, each with its unique strengths and applications.

The generative models, such as the Generative Translated Branch and Generative Direct Compound Branch, show significant potential in producing highly realistic animations by leveraging phonemes and multimodal inputs. Notable examples include the works by Taylor et al. and Facebook's Codec Avatars, which demonstrate impressive results in speaker-independent animations and full-face motion, respectively.

On the other hand, the adaptive models, like those in the Adaptive Specialist Branch and Adaptive Generalist Branch, focus on altering pre-existing videos to achieve lip synchronization. These approaches, exemplified by Suwajanakorn et al.'s Obama model and Prajwal et al.'s LipGAN, offer compelling solutions for specific characters or generalist applications, albeit often requiring larger and more diverse datasets.

The introduction of a new taxonomy to classify these methods based on logistical aspects rather than model structures or technological details provides a clearer framework for understanding and comparing the various techniques. This taxonomy aids in identifying the most suitable approach for different applications, from virtual assistants and gaming dialogues to telepresence and multimedia localization.

Moreover, the research underscores the importance of addressing challenges such as model training cost, dataset availability, and silent moment distributions in audio data. Innovative solutions, such as the metric for evaluating audio quality concerning silent points and the use of external voice decoders for phoneme translation, enhance the performance and realism of the generated animations.

Overall, this comprehensive study advances the field of audio-visual synchronization by offering a robust classification system, identifying key challenges, and showcasing state-of-the-art techniques. The proposed models and methodologies pave the way for future developments, enhancing the capabilities and applications of virtual assistants, gaming, and other interactive digital media.

%=============================================================================
\bibliographystyle{ieeetr}
\bibliography{ETAPA3_ET_5_reduzido}

\end{document}